\documentstyle[proc]{rspublic}
\begin{document}
\title[Exact General Solutions to Extraordinary N-body Problems]{Exact
General Solutions to Extraordinary N-body Problems}
\author[D. Lynden-Bell and R.M. Lynden-Bell]{D. Lynden-Bell \& R.M. Lynden-Bell} 
\affiliation{School of Mathematics \& Physics, The Queen's University,
Belfast  BT7 1NN, UK }

\maketitle

\label{firstpage}

\begin{abstract}

\noindent
The N-body problem in which the force on the $I{\rm th}$ body is \break
${\bf F}_I = k \sum_{J} m_Im_J ({\bf x}_J - {\bf
x}_I)/r^3$ where $r$ is the time dependent radius of the whole system
is solved exactly for all initial conditions.  Here, writing ${\overline {\bf x}}$ for the position of the barycentre,  $r^2 = M^{-1}
\sum_I m_I ({\bf x}_I - {\overline {\bf x}} )^2$.

Also solved are all N-body problems for which the potential energy is
of the form $V = V(r)$ for any function $V$ where $r$ is given above.
The first problem is the special case $V \propto r^{-1}$.  In all cases
$r$ vibrates for ever, just like the radius to a body which has angular
momentum about a centre of attraction due to the potential $V(r)$.
Violent Relaxation does not occur in these systems.  When $r$ vibrates
as in a highly eccentric orbit the exact solution should prove to be a
useful test of N-body codes but a severer test follows.

We show further that when ${\bf F}_I$ is supplemented by an inverse
cubic repulsion between each pair of bodies $- k' {\sum_{J
\neq I}} m_I m_J ({\bf x}_J - {\bf x}_I)/(x_J - x_I)^4$ there is still
no damping of the fundamental breathing oscillation which performs
its non-linear periodic motion for ever.  This is a special case of
continually breathing systems with potentials $V_0(r) + r^{-2} V_2
\left( ({\bf x}_I - {\bf x}_J) /r \right )$.

A remarkable new statistical equilibrium is found for systems whose
size is continually changing!

\medskip

\indexsize{\centerline{Keywords: N-body systems; Newton; orbits; dynamical systems;} 
\centerline{statistical mechanics; pulsating equilibria}}

\end{abstract}

\section{Introduction}
In 1995 one of us (Lynden-Bell, R.M. 1995, 1996) showed that the
classical statistical mechanics of systems of N bodies interacting
through mutual potential energies of the form $$V = N F (\sum
{\bf q}^2_I/N) \eqno (1.1)$$
could be exactly calculated for any chosen function $F$.  By choice of
a suitable function $F$, she found a case with a simple phase
transition which was calculated for any $N$.  

When the particles are of equal mass and when ${\bf q}_I$ measures the
displacement of the $I^{th}$ particle from the barycentre ${\overline
{\bf x}} , {\sum\limits_I}{\bf q}^2_I/N$ may be rewritten as a mutual
potential energy
$$N^{-1} {\bf \sum}\, {\bf q}^2_I = N^{-1} {\sum_I} ({\bf x}_I
- {\overline {\bf x}})^2 = \sum_{\ \ I \ <} \sum_{\! J} ({\bf x}_I - {\bf
x}_J)^2/N^2 \ . \eqno (1.2)$$ 

More than 300 years previously Newton showed that the N-body problem
in which the force on body $I$ due to body $J$ was $$k \, m_I m_J
({\bf x}_J - {\bf x}_I)$$ could be exactly solved with all the
interactions present (Newton (1687), see also Chandrasekhar (1995)).
Newton's case corresponds to a potential energy
$$V = {\textstyle {1 \over 2}} k \sum_{\ \ I \ <} \sum_{\! J} m_I m_J ({\bf x}_I
- {\bf x}_J)^2 \, . \eqno (1.3)$$
When the masses are equal the expression (1.3) is the linear-$F$ case of
expression (1.1) by virtue of (1.2).  As the statistical mechanics
associated with expression (1.1) is still beautiful even when $F$ is any
non-linear function, we are led to wonder whether Newton's exact
solution to the N-body problem could be generalised to cover potential
energies of the form $V = V(r)$.  Here $V$ is any function and $r$ is
the mass weighted root mean square size of the system 
$$r^2 = M^{-2} \sum_{\ \ I \ <} \sum_{\! J} m_I m_J ({\bf x}_I - {\bf
x}_J)^2 = M^{-1} {\sum _I} m_I ({\bf x}_I - {\overline {\bf
x}})^2 \eqno (1.4)$$
where
$$M = \sum m_I \ \ {\rm and} \ \ {\overline {\bf x}} = M^{-1}
{\sum_I} m_I {\bf x}_I \ . \eqno (1.5)$$

Following this idea we found a beautiful dynamics in 3$N$ dimensional
space which precisely mimics normal orbit theory in 3 dimensions.

\section{Solutions to the N-body Problems}

The equations of motion are

$$m_I {\ddot {\bf x}}_I = - \partial V/\partial {\bf x}_I \, . \eqno
(2.1)$$ Provided $V$ is a mutual potential energy, i.e., a function of
the ${\bf x}_I - {\bf x}_J$ then ${\sum\limits_I} \partial
V/\partial {\bf x}_I =0$ so ${\ddot {\overline {\bf x}}} = 0$ and
$${\overline {\bf x}} = {\overline {\bf x}}_0 + {\bf u}t \eqno (2.2)$$

Writing ${\bf q}_I = {\bf x}_I - {\overline {\bf x}}$ we have $r^2 =
M^{-1} {\sum\limits_I} m_I {\bf q}^2_I$ so in centre of mass coordinates the
equations of motion for the ${\bf q}_I$ are
$$m_I {\ddot {\bf q}}_I = - {\partial V \over \partial {\bf q}_I} = -
{\partial V \over \partial r} {m_I \over M} {{\bf q}_I \over r} \ . \eqno
(2.3)$$
We now invent a 3$N$ dimensional vector ${\bf r}$ of length $r$, whose
components are
$${\bf r} = \left( {\sqrt {m_1 \over M}}\, {\bf q}_1 \, , \, {\sqrt
{m_2 \over M}}\, {\bf q}_2 \ \ldots \ {\sqrt {m_I \over M}} \, {\bf
q}_I \ \ldots \
\right) \, . \eqno (2.4)$$
The first 3 components of ${\bf r}$ involve the position of particle 1
and the next 3 particle 2, etc.  Multiplying (2.3) by $\sqrt{M/m_I}$ we
see that those equations for all $I$ can be rewritten using unit vector ${\hat {\bf r}}$
$$M {\ddot {\bf r}} = - {\partial V \over \partial r} \, {{\bf r}
\over r} = - V' {\hat {\bf r}}\ . \eqno (2.5)$$ 
For insight please consult here the note added in proof.

Let $\alpha$ and $\beta$ be indices that run from 1 to 3$N$.  Then
taking the antisymmetric product of (2.5) with ${\bf r}$ we find
$$M {d \over dt} \left ( r_\alpha {\dot r}_\beta - r_\beta {\dot
r}_\alpha \right) = 0$$
so the hyper-angular-momenta (per unit mass) are all conserved
$$r_\alpha {\dot r}_\beta - r_\beta {\dot r}_\alpha = L_{\alpha \beta}
= \, {\rm const} \, = -L_{\beta \alpha} \ . \eqno (2.6)$$

These quantities are not just normal angular momenta.  They involve
among others the $x$ component of particle 1 multiplied by the rate of
change of the $x$ component of particle 2, i.e., $x_1 {\dot x}_2$ and
also such terms as $x_1 {\dot y}_2$, etc.  Only when both $\alpha$ and
$\beta$ are in the range 1, 2, 3 does $L_{\alpha \beta}$ reduce to
components of the normal angular momentum of particle 1 (in that
case).  The trace of the square of the antisymmetric tensor $L_{\alpha
\beta}$ is
$$L_{\alpha \beta} L_{\alpha \beta} = 2 \left[ r^2 {\dot {\bf r}}^2 -
({\bf r} \cdot {\dot {\bf r}})^2 \right] = 2 L^2 \ . \eqno (2.7)$$
Notice that if we were in 3 dimensions $L^2$ would be just $({\bf r}
\times {\dot {\bf r}})^2$ so it would then be the square of the
specific angular momentum.  We shall refer to $ML$ as the total
hyper-angular-momentum and $ML_{\alpha \beta}$ as the
hyper-angular-momentum.  Taking the inner product of equation (2.5)
with ${\dot {\bf r}}$ and integrating, we obtain the energy equation
$${\textstyle {1 \over 2}} M {\dot {\bf r}}^2 + V(r) = E \, . \eqno
(2.8)$$
Now $${\dot r} = {d \over dt} ({\bf r} \cdot {\bf r})^{1/2} = {\bf r}
\cdot {\dot {\bf r}}/r \eqno (2.9)$$
so from (2.7) $${\dot {\bf r}}^2 = L^2 r^{-2} + {\dot r}^2 \eqno (2.10)$$
and (2.8) may now be rewritten $${\textstyle {1 \over 2}} M ({\dot r}^2
+ L^2 r^{-2}) + V = E \ . \eqno (2.11)$$
Differentiating (2.11) and dividing by ${\dot r}$ we get the
equation of motion for scalar $r$
$$M({\ddot r} - L^2 r^{-3}) = -V'(r) \ . \eqno (2.12)$$
This we recognise as the equation of motion of a particle of mass $M$
and angular momentum $ML$ in the central potential $V (r)$.  Equation (2.11) is
of course its energy equation and it may be integrated in the form
$$t - t_0 = \int ^r {dr \over \sqrt{2M^{-1} (E - V(r)) - L^2 r^{-2}}} \ ,
\eqno (2.13)$$ which gives the standard relation for the
periodic{\footnote {Periodic when the system is bound.}} function
$r(t)$.  Equation (2.13) integrates nicely for the standard cases $V \propto  1/r$,
$V \propto r^2$ and the isochrone $V \propto (b + s)^{-1}$ where $s^2 =
r^2 + b^2$.  These potentials may be supplemented by an extra term in
$r^{-2}$ giving a $r^{-3}$ supplementary force but they comprise the
only central orbits that can be solved involving nothing more
complicated than trigonometric functions (Eggen et al. 1962, Evans et
al. 1990).  There are many others that involve elliptic functions,
etc., but for them the expression (2.13) is as simple as the function.
We shall return to (2.13) later but for the present we merely note that
it determines the periodic$\dagger$ function $r(t)$.

By analogy with normal central orbits the form of equation (2.12)
strongly suggests that we invent an angle $\phi$ such that
$$r^2 {\dot {\phi}} = L \ . \eqno (2.14)$$
Then from (2.11) we eliminate $t$ and integrate to find
$$\phi = \int ^r _{r_{\min}} {L \ dr \over r^2 \sqrt{2M^{-1} [E + V(r)]
- L^2r^{-2}}}\ , \eqno (2.15)$$ just as in ordinary orbit theory.  We
define $\phi$ by (2.15) and so there is no constant of integration.  It
gives $\phi$ as a function of $r$, or $r$ as a periodic function of
$\phi$.  Thus we have determined scalar $r$ but we need to find all
components of the 3$N$ vector ${\bf r}$ to determine the motion
completely.  Although $V'(r)/r$ may now be regarded as a known
function of $t$ nevertheless the form of (2.5) does not at first look
very encouraging, except for Newton's case for which $V = {\textstyle {1 \over 2}}kM^2r^2$ so
$V'(r)/r$ is constant.  However, a wonderful simplicity will emerge for
the general case too.  For a moment we discuss Newton's case.
Equations limited to this case will have the letter N appended to the
equation number.  Evidently (2.5) becomes
$${\ddot {\bf r}} = - k M {\bf r} \eqno (2.16{\rm N})$$
so each component of ${\bf r}$ vibrates harmonically, all with angular
frequency $\sqrt{kM}$.  Thus each body orbits in an ellipse centred on
the barycentre and moving with it and the whole motion relative to the
barycentre is periodic with period $2 \pi / \sqrt{kM}$.  Newton
obtained this solution by realising that the linear law of attraction,
when suitably mass weighted, gave a net total force on each particle
due to all the others which was directed toward the barycentre and in
constant proportion to distance from it.

A similar simplicity emerges in the general case when we solve for
${\hat {\bf r}} = {\bf r}/r$ as a function of $\phi$.  To do this we write 
$$
{\ddot {\bf r}} = {d \over dt} \left ( r{\dot {\hat {\bf r}}} + {\dot
r} {\hat {\bf r}} \right ) = {d\over dt} \left ( {L \over r} {d {\hat
{\bf r}} \over d \phi} + {\dot r} {\hat {\bf r}} \right) = {L^2 \over
r^3} \, {d^2 {\hat {\bf r}} \over d \phi^2} + \ddot r {\hat {\bf r}} \ ,
\eqno (2.17)$$
where two terms cancelled at the last step.  Now scalar $\ddot r$ is
given by (2.12) so, inserting (2.17) into (2.5) and multiplying by $r^3
L^{-2}$, we find the lovely result
$$d^2 {\hat {\bf r}}/d \phi^2 + {\hat {\bf r}} = 0 \ . \eqno (2.18)$$
Thus the unit vector ${\hat {\bf r}}$ vibrates harmonically in $\phi$
with period $2 \pi$.  For Newton's case, and for the hyper-Keplerian
case $V \propto 1/r$, the magnitude $r$ is also periodic with period $2
\pi$ in $\phi$.  But more generally, as in the isochrone, $r$ is
periodic but with a $\phi$ period that is incommensurable with $2 \pi$
(in general).  The general solution of (2.18) is given by
$${\hat {\bf r}} = {\bf A} \sin \phi + {\bf B} \cos \phi \eqno (2.19)$$
where ${\bf A}$ and ${\bf B}$ are 3$N$-vector constants of
integration.  But ${\hat {\bf r}}$ is a unit vector so
$$\displaylines {1 = {\bf A}^2 \sin^2 \phi + 2{\bf A} \cdot {\bf B}
\sin \phi \cos \phi + {\bf B}^2 \cos^2 \phi = {\textstyle {1 \over 2}}
({\bf A}^2 + {\bf B}^2) - \cr - {\textstyle {1 \over 2}} ({\bf A}^2 -
{\bf B}^2) \cos 2 \phi + {\bf A} \cdot {\bf B} \sin 2 \phi \ .} $$
Since this must be true for all $\phi$ we find
$${\bf A}^2 = {\bf B}^2 = 1 \ \ {\rm and} \ \ {\bf A} \cdot {\bf B}
= 0 \ , \eqno (2.20)$$
so ${\bf A}$ and ${\bf B}$ must be orthogonal unit 3$N$-vectors.
There is a further restriction on these constants; because the ${\bf
q}_I$ are the centre of mass coordinates they must obey
$$0 = \left( {\sum _I} m_I {\bf q}_I \right) _j = \left( {\sum_I} \sqrt{m_I}\, r_{3I-3+j} \right) \sqrt{M}\, r \eqno {\rm for}
\ j=1,2,3$$
therefore,
$$0 = {\sum _I} \sqrt{m_I} \, A_{3I-3+j}$$
\vspace{-.2in}

\hfill(2.21)
\vspace{-.1in}

\noindent
and 
\vspace{-.1in}
$$0 = {\sum_I} \sqrt{m_I} \, B_{3I-3+j}$$

To construct constants ${\bf A}$ and ${\bf B}$ satisfying the
constraints (2.20) and (2.21) we proceed as follows:

Take $m_1$ to be the largest mass; choose ${\overline B}_\alpha \ \,
\alpha >3$ and ${\overline A}_\alpha \ \, \alpha > 3$ arbitrarily, but
set ${\overline A}_j = {\sum \limits_{I>1}} \sqrt{{m_I \over m_1}} \,
{\overline A}_{3I-3+j}$ for $j = 1,2,3$ and a similar relation for
${\overline B}_j$.  Then ${\overline A}_j$ and ${\overline B}_j$
satisfy (2.21) and so will $\lambda {\overline {\bf A}}$ and $\mu
({\overline {\bf B}} - \nu {\overline {\bf A}})$ where $\lambda$, $\mu$
and $\nu$ are any scalars, since (2.21) are linear.  

Now choose $\lambda = |{\overline {\bf A}}|^{-1}$ and set ${\bf A} =
\lambda {\overline {\bf A}}$ so that ${\bf A}$ is a unit vector;
furthermore set ${\bf B}' = {\overline {\bf B}} - ({\overline {\bf B}}
\cdot {\bf A}) {\bf A}$ so that ${\bf B}' \cdot {\bf A} = 0$;
finally normalise by writing ${\bf B} \equiv \mu {\bf B}' $ with $\mu
= {\textstyle {1 \over |{\bf B}'|}}$.  Then ${\bf A}$ and ${\bf B}$ so
constructed are the general unit vectors satisfying the constraints
(2.20) and (2.21).  Thus our general solution for ${\bf r}$ is
$${\bf r} = r {\hat {\bf r}} = r ({\bf A} \sin \phi + {\bf B} \cos
\phi) \eqno (2.22)$$
Thus relative to the centre of mass every particle's orbit becomes a centred ellipse after a universal scaling by $r^{-1}$.  This rescaling factor is time-dependent and $r$ behaves like the radius to a particle in a central orbit which is an eccentric ellipse in the Keplerian case.  This is our main result.

Differentiating (2.22) with respect to $t$ one finds
that (using $r^2 \dot \phi = L)$
$$L_{\alpha \beta} \equiv  r_\alpha \dot r_\beta - r_\beta \dot
r_\alpha = (B_\alpha A_\beta - A_\alpha B_\beta)L \ . \eqno (2.23)$$

It is of interest here to count constants of integration.  ${\bf A}$
and ${\bf B}$ together have 6$N$ components but (2.20) and (2.21) give nine
constraints, thus there are 6$N$-9 freedoms so far.  In the solutions
for $r(t)$ and $r(\phi)$ there are three further constants $L, E$ and
$t_0$.  Thus $r(t)$ depends on 6$N$-6 constants.  To these we may add
the constants ${\overline {\bf x}}_0$ and ${\bf u}$ involved in the
motion of the centre of mass, and our final solution depends on 6$N$
arbitrary constants.  This equals the number of initial positions and
velocities that determine the motion, which verifies that we have the
general solution.  

The solution is only fully explicit when we can carry out the
integrations (2.13) and (2.15) which determine $r$ and $\phi$.  We do not
need this for Newton's case because we already gave the solution.
When $V \propto r^{-1}$, the hyper-Keplerian case, it is actually
simpler to follow Hamilton's treatment of the problem with his
eccentricity vector later used by Runge in 1919 and Lenz in 1924 (see
Hamilton 1845, Laplace 1799, Goldstein 1976, or Chandrasekhar
1995).  This was certainly known to Laplace and Bernoulli and can be
traced to Newton.  The method avoids the awkwardness of having to
treat the elliptic, parabolic and hyperbolic cases separately.  With
$V = -k M^2/r$ equation (2.5) becomes
$${\ddot {\bf r}} = -kM {\hat {\bf r}}/r^2 \eqno (2.24{\rm K})$$
contracting with $L_{\alpha \beta}$ we find
\begin{eqnarray*}
{d \over dt} (L_{\alpha \beta} {\dot r}_\beta) & = & - kM \left( {\hat
r}_\alpha {{\dot r}_\beta \over r} {\hat r}_\beta - {\hat r}_\beta
{{\dot r}_\alpha \over r} {\hat r}_\beta \right) \\ & = & kM {d \over
dt} \left( {r_\alpha \over r} \right) \ .  
\end{eqnarray*}
Hence
$$L_{\alpha \beta} {\dot r}_\beta = kM ({\hat r}_\alpha + e_\alpha)
\eqno (2.25{\rm K})$$
where ${\bf e}$ is a 3$N$-vector eccentricity.  Contracting with
${\bf r}_\alpha/r$ and using the antisymmetry of $L_{\alpha \beta}$ we find, 
$$\ell/r = 1 + {\bf e} \cdot {\hat {\bf r}} \ ,  \eqno (2.26{\rm
K})$$
where $$\ell = L^2/(kM) \ . \eqno (2.27{\rm K})$$

If $e$ is the magnitude of ${\bf e}$ and $\phi$ is the angle between
${\bf e}$ and ${\hat {\bf r}}$, then this equation becomes the equation
of a conic section in the $r, \phi$ plane, i.e.,
$$\ell/r = 1 + e \cos \phi \ . \eqno (2.28{\rm K})$$

To demonstrate that $\phi$ is indeed the angle called by that name
previously, we first square (2.25K) writing out $L_{\alpha \beta}$ in
full and using (2.7) on the left and eliminating ${\bf e} \cdot {\hat
{\bf r}}$ via (2.26K) on the right.  This gives just as for the 3
dimensional case
$$L^2 {\dot {\bf r}}^2 = k^2 M^2 (e^2-1) +2 L^2Mk r^{-1} \ ,$$
which we rewrite in the form of the energy equation (2.8):
$$
{\textstyle {1 \over 2}} M {\dot {\bf r}}^2 - {kM^2 \over r} = {\textstyle {1 \over 2}} {k^2M^3
(e^2 - 1) \over L^2} = E \ . \eqno (2.29{\rm K})$$
Remembering (2.10) we have
$$
\dot r^2 = 2 \left ( {E \over M} + k {M \over r} \right) - {L^2 \over
r^2} = {k^2M^2 \over L^2} \left[ e^2 - \left( {\ell \over r} - 1
\right)^2 \right] \  . \eqno (2.30{\rm K})$$
Differentiating (2.28K)
$$
{L^2 \dot r \over kM} = e \sin \phi \, (r^2 \dot \phi) \ , \eqno (2.31{\rm
K})$$
but by (2.30K) and (2.28K) $$\dot r = {kM \over L}\, e \sin \phi \ . \eqno
(2.32{\rm K})$$
Thus $r^2 \dot \phi = L$ and our new $\phi$ can only differ from our
old one by a constant $\chi$.  Such a constant is irrelevant as it
merely makes a transformation 
\begin{eqnarray*}
{\bf A} & \rightarrow & {\bf A} \cos \chi - {\bf B} \sin
\chi \\ {\bf B} & \rightarrow & {\bf B} \cos \chi + {\bf A}
\sin \chi 
\end{eqnarray*} 
on our constants of the motion.  The constraints (2.20) and (2.21) are
invariant to such transformations.  Thus our orbit in 3$N$ space is
given by
$${\bf r} = {\ell \over 1 + e \cos \phi} \, ({\bf A} \sin \phi + {\bf
B} \cos \phi) \ . \eqno (2.33{\rm K})$$
This is our prime result and gives the ${\bf x}_I$ via (2.4) and (2.2).  Relative to the centre of mass each particle's orbit lies in a plane and eliminating $\phi$ in favour of coordinates $x$ and $y$ in that plane we find it is a conic section.  When $e<1$ the ellipse has in general neither its focus nor its centre at the centre of mass.  This is clear from the geometrical interpretation given in the note added in proof.

The relationships between $r, \phi$ and $t$ are given as usual by
Kepler's equation which one may derive from (2.13) and from (2.28K) and
(2.31K):
$$dt = {dr \over {\dot r}} = {d \left( {\ell \over 1 + e \cos \phi} \right)
\over {kM \over L} e \sin \phi} = {L^3 \over k^2M^2} \, {d \phi \over
(1 + e \cos \phi)^2} \ . \eqno (2.34{\rm K})$$

Defining $a = GM/(-2 \epsilon)$ these give, writing $r = a(1 - e \cos
\eta)$, 
$$t - t_{\min} = GM(-2 \epsilon)^{-3/2}\, (\eta - e \sin \eta) \ , \eqno
(2.35{\rm K})$$
and 
$$\phi = 2 \tan^{-1} \left( \sqrt{{1-e \over 1+e}} \tan {\eta \over 2}
\right) \ , \eqno (2.36{\rm K})$$ where $\phi$ is measured from
pericentre.  From this one deduces $\cos \phi = {\cos \eta - e \over 1
- e \cos \eta}$ and $\ell = a (1 - e^2)$.  The generalisations of
these formulae for the isochrone potential are given in the Appendix.

\section{Generalisation to Breathing Systems and their new Statistical
Mechanics}

When the potential is of the form\footnote{$\bf r$ is constrained by
the barycentre constraint.  It is understood that $V_2$ takes values
dependent only on the subspace to which ${\hat {\bf r}}$ is
constrained} $V_0 (r) + r^{-2} V_2 ({\hat {\bf r}})$ equation (2.5)
becomes
$$M {\ddot {\bf r}} = - V_0' (r) {\hat {\bf r}} + 2 r^{-3} V_2 {\hat
{\bf r}} - r^{-3} \partial V_2/d {\hat {\bf r}} \ ,  \eqno (3.1)$$
where the last term in the force is not hyperradial.  Indeed since
$V_2$ is independent of $r$ it is constant along each radial line so
its gradient is automatically perpendicular to ${\bf r}$.  We
therefore form a Virial theorem by taking the dot product of (3.1)
with ${\bf r}$ and using
$${\bf r} \cdot {\ddot {\bf r}} = {d \over dt} ({\bf r} \cdot {\dot
{\bf r}}) - {\dot {\bf r}}^2 = {\textstyle {1 \over 2}} {d^2 \over dt^2} (r^2) -
{\dot {\bf r}}^2 \ . \eqno (3.2)$$
We thus obtain writing $T = \sum {\textstyle {1 \over 2}} M {\dot {\bf
r}}^2$
$${\textstyle {1 \over 2}} M {d^2 \over dt^2} (r^2) = 2T - r V_0' + 2 r^{-2} V_2 \
. \eqno (3.3)$$
But by energy conservation $$T + V_0 + r^{-2} V_2 = E \ ,  \eqno (3.4)$$ so
$${\textstyle {1 \over 2}} M {d^2 \over dt^2} (r^2) = 2 E - {1 \over r} {d \over
dr} (r^2 V_0) \ . \eqno (3.5)$$
Notice that $V_2$ has disappeared from this equation which is now an
equation for the scalar $r$ only.  This separation of the motion of
the scaling coordinate $r$ from the rest of the dynamics only occurs
when the potential is of the special form we have chosen with the
`angularly' dependent part scaling as $r^{-2}$.  Multiplying (3.5) by
$dr^2/dt$ and integrating
$${\textstyle {1 \over 4}} M \left( {dr^2 \over dt} \right)^2 = 2 E r^2 - 2 r^2 V_0
- M {\cal L}^2 \ , \eqno (3.6)$$
where the final term is the constant of integration and ${\cal L}$ has
the same dimensions as our former $L$ whose part it plays.  Dividing
by $2r^2$ we now have
$${\textstyle {1 \over 2}} M ({\dot r}^2 + {\cal L}^2 r^{-2}) + V_0 (r) = E \
. \eqno (3.7)$$
The equation of motion for $r$ follows by differentiation
$$M({\ddot r} - {\cal L}^2 r^{-3}) = - \partial V_0/\partial r \
. \eqno (3.8)$$
Thus $r$ vibrates as though it were in a central orbit of a particle
of mass $M$ and angular momentum $M {\cal L}$ with a central potential
$V_0(r)$.  Integrating (3.7) we see that for bound orbits $r$ vibrates
for ever with a period
$$P = \oint {dr \over {\sqrt{2M^{-1} (E - V_0 (r)) - {\cal L}^2
r^{-2}}}} \ . \eqno (3.9)$$

In the above we have assumed that ${\cal L}^2$ is positive and that
$V_0 (r)$ is not so strongly attractive that it beats the centrifugal
repulsion.  Then for bound states there are normally simple zeros of
the expression in the surd above.  When the constant ${\cal L}^2$ is
negative, one may change the nomenclature to make it positive by adding
a constant $K$ to $V_2$ and subtracting $K/r^2$ from $V_0$ so as to
leave $V$ unchanged.  While this makes ${\cal L}^2$ positive it may
make $V_0$ so attractive that $r$ can reach the origin.  If we again
define an angle $\phi$ such that $r^2 {\dot \phi} = {\cal L}$ and plot
an orbit for $r$ in $r, \phi$ space, then such orbits approach the
origin in finite time following logarithmic spirals (for small $r$).
They re-emerge with $\phi$ still increasing but with $r$ now
increasing on logarithmic spirals.  Although the time near the origin
is finite, the increment in $\phi$ on passing through the origin is
infinite.  The outside turning points of these orbits will be just as
for others.

The form of the potential energy in (3.4), $V_0 (r) + r^{-2} V_2 ({\hat
{\bf r}})$ is the generalisation to $3N$ dimensions of the 3
dimensional potential $V_0 + r^{-2} V_2 (\theta, \phi)$ for which the
radial motion separates from the transverse through the exact integral
${\textstyle {1 \over 2}} m {\bf h}^2 + V_2 (\theta, \phi) =$ const
where ${\bf h} = {\bf r} \times {\bf v}$.  Further separations occur
if $V_2 = U (\theta) + W_\phi (\phi)/ \sin^2 \theta$.  Still further
integrals can occur for special $V_0$ such as the simple harmonic and
the Kepler potential, see Evans, 1990.  In $3N$--3 dimensions each
separable potential gives an exactly soluble N-body problem.  Marshall
\& Wojciechowski show that the general case occurs in hyperellipsoidal
coordinates in $3N$--3 dimensions and that other cases, such as the
hyperspherically separable one we have used, can be found as
degenerate cases of hyperellipsoidal coordinates.

We now apply the general theory of this section to the case mentioned
in the abstract for which
$$V_0 = {\textstyle {1 \over 2}} k \sum_{\ \ I \ <} \sum_{\! J} m_I
m_J ({\bf x}_I - {\bf x}_J)^2 = {\textstyle {1 \over 2}} kM^2 r^2 \ ,
\eqno (3.10)$$ and
$$r^{-2} V_2 = {\textstyle {1 \over 2}} k' \sum_{\ \ I \ <} \sum_{\!
J} m_I m_J ({\bf x}_I - {\bf x}_J)^{-2} \ , \eqno (3.11)$$ with $k$ and
$k'$ positive.  Then we have the system of N bodies which attract each
other linearly according to their separations and repel each other
with an inverse cubic repulsion.  Since the scale $r$ of ${\bf x}_I -
{\bf x}_J$ cancels out in $V_2$ we see that indeed $V_2 = V_2 ({\hat
{\bf r}})$.  Thus the general theory applies and the period of $r$ is
given by (3.9) which gives
$$P = {\pi \over {\sqrt{ kM}}} \  . \eqno (3.12)$$
This is only half the period in $\phi$ because for a central ellipse
$r$ undergoes two oscillations as $\phi$ increases by $2 \pi$.

Thus the radius of such a system will continue to vibrate for ever and
shows no Violent Relaxation (Lynden-Bell, 1967).  With both attractive
and repulsive forces present, it may be possible to get a giant
lattice solution but the increased pressures near the centre will give
interesting radial distortions to any such lattice as in a planet.

Even though the above hyper-radial motions are simple the ${\hat {\bf
r}}$ motions are not, may show relaxation phenomena and we find below
that they may even achieve a slightly modified form of equilibrium
even though the scaling variable $r$ continues to vibrate at large
amplitude!

To see this we first return to the general case with $V_0 (r)$ and
$V_2 ({\hat {\bf r}})$ any functions and look for the equations of
motion of the ${\hat {\bf r}}$.  These follow from (3.1) if we
remember that $${\ddot {\bf r}} = {d \over dt} (r {\dot {\hat {\bf r}}}
+ \dot r {\hat {\bf r}}) = \ddot r {\hat {\bf r}} + {1 \over r}
{d \over dt} (r^2 {\dot {\hat {\bf r}}}) \ ; $$ substituting this into (3.1)
and eliminating ${\ddot r}$ via (3.8) yields 
$${M r^2} {d \over dt} \left( { {d {\hat {\bf r}}
\over dt}} \right) = \left( M {\cal L}^2 + 2 V_2 \right) {\hat {\bf r}}
- {\partial V_2 \over \partial {\hat {\bf r}}} \ . $$

Putting $r^{-2} d/dt = d/d \tau$ this may be rewritten as an
autonomous equation for ${\hat {\bf r}} (\tau )$
$$M d^2 {\hat{\bf r}}/d \tau ^2 = -M {\cal L}^2 {\hat {\bf r}} + 2 V_2
{\hat {\bf r}} - \partial V_2 / \partial {\hat {\bf r}} \ . \eqno
(3.13)$$
The radial component of this equation is irrelevant; it follows from
the condition $| {\hat {\bf r}}| = 1$ and the transverse components.
We now write ${\bf Q}_I = \sqrt{m_I/M}\, {\bf q}_I /r$ so that the ${\bf
Q}_I$ are the components of ${\hat {\bf r}}$ taken in triples.  Then
the centre of mass constraint becomes ${\sum_I} \sqrt{m_I}\, {\bf
Q}_I = 0$.  When we expressed $V_2$ in terms of our original ${\bf
x}_I$ it depended only on differences so $V_2 ({\bf x}_1 + {\bf
\Delta} , \ldots {\bf x}_I + {\bf \Delta} \ldots x_{\bf N} + {\bf
\Delta})$ was independent of ${\bf \Delta}$.  This means that when we
write $V_2$ as a function of our new coordinates ${\bf Q}_I$ it has
the property that $$V_2 \left( {\bf Q}_1 + \sqrt{m_1 \over M} {{\bf \Delta}
\over r} , \ldots {\bf Q}_N + \sqrt{{m_N \over M}} {{\bf \Delta} \over
r} \right)$$ is independent of ${\bf \Delta}$.  Differentiating with
respect of $\Delta$ and then setting ${\bf \Delta} = 0$ we deduce that
$${\sum_I} \sqrt{m_I}\, \partial V_2 / \partial {\bf Q}_I = 0 \
. \eqno (3.14)$$

Now consider the Lagrangian $\L ({\bf Q}_I' , {\bf Q}_I)$ where ${\bf
Q}_I' = d {\bf Q}_I / d \tau$ and $$\L = {\sum_I} {\textstyle {1 \over
2}} M {\bf Q}'^2_I - V_2 ({\bf Q}_1 \ldots {\bf Q}_N) \ , \eqno (3.15)$$
and the ${\bf Q}_I$ are subject to the constraints
$${\sum_I} {\bf Q}^{2}_I = 1 \ \ {\rm and} \ \ {\sum_I} \sqrt{m_I} \, 
{\bf Q}_I = 0 \ . \eqno (3.16)$$
Consider also the variational principle in $\tau$-time $\delta \int \L
({\bf Q}'_I , {\bf Q}_I) d \tau = 0$.

Lagrange's equations are
$$M {\bf Q}''_I = - \partial V_2 / \partial {\bf Q}_I + \lambda (\tau) {\bf
Q}_I + {\mbox {\boldmath $\mu$}} (\tau) \sqrt{m_I} \ , \eqno (3.17)$$
where $\lambda$ and ${\mbox {\boldmath $\mu$}}$ are Lagrange
multipliers corresponding to the constraints (applied for all
$\tau$).  Multiplying (3.17) by ${\sqrt{m_I}}$ and summing using
(3.14), the second constraint equation (3.15) gives $
{\mbox {\boldmath $\mu$}} \equiv 0$.  Evidently (3.17) and (3.13) are
equivalent since ${\hat {\bf r}} = ({\bf Q}_1, {\bf Q}_2 \ldots {\bf
Q}_N)$.  So (3.13) follows from the Lagrangian $\L$.  The `energy'
equation for the ${\hat {\bf r}}$ follows from $r^2 (3.4) - {\textstyle{1 \over 2}} (3.6)$
when we use ${\dot {\bf r}}^2 = r^2 (d {\hat {\bf r}}/dt)^2 + {\dot
r}^2$ and we get
$${\textstyle {1 \over 2}}M \left( {d {\hat {\bf r}} \over d \tau} \right) ^2 + V_2
({\hat {\bf r}}) = {\textstyle {1 \over 2}} M {\cal L}^2 \ , \eqno (3.18)$$
where by definition of ${\bf Q}_I$
$$\left( d {\hat {\bf r}} \over d \tau \right) ^2 = {\sum_I}
{\bf Q}'^2_I \ . $$

In the statistical mechanics that follows we shall be concerned with
the case in which $N$ is large so that there are many terms in this
sum.

\subsection{Statistical Equilibrium of a system in large amplitude
oscillation} 

In one sense it is surprising that a system may achieve a statistical
equilibrium while its scale continues to oscillate (or in the unbound
case evolve).  However this only occurs for these special systems in
which the large scale oscillations separate from the rest of the
dynamics and only then when the potential $V_2$ is sufficiently
complicated to give quasi-ergodic motion in ${\hat {\bf r}}$.  Once
the $r$ motion has been separated off, we have shown that the motion of
${\bf Q}_I$ follows from a Lagrangian in $\tau$-time and that $V_2$
can be any function of the normalised relative coordinates, so it can
achieve the necessary complications.  Since the ${\bf Q}_I$ motion
pays no attention to the $r$ motion there is no reason why it should
not achieve such an equilibrium.  Whether such an equilibrium is a
thermal equilibrium or not depends on semantics.  The quantity shared
by the different ${\bf Q}_I$ is not energy but hyper-angular-momentum,
$M {\cal L}^2$, and the time is replaced by $\tau$ but, as we shall see,
the equilibrium remains at all phases of the oscillation and does not
depend on the timescale of the relaxation as compared with the period
of oscillation.  In many respects it is a true generalisation of the
usual concept of statistical equilibrium albeit with sufficient
differences to  make it interesting.  We have $N, {\bf Q}_I$ subject
to the 4 constraints
$$\sum \sqrt{m_I} {\bf Q}_I = 0 \eqno (3.19)$$
$$\sum {\bf Q}_I^2 = 1 \ , \eqno (3.20)$$
and having equations of motion following the Lagrangian (3.15) and
having `energy' given by (3.18).  When there are very many particles
present the fixed centre of mass constraints (3.19) are unimportant.
They only give changes of order $1/N$ in the results and are in any
case statistically satisfied by the equilibrium found without imposing
them.  As they significantly complicate the arguments while adding
very little to the result we shall now solve the problem when only the
constraint (3.20) is imposed and we shall specialise to an interaction
$V_2$ equivalent to a hard sphere gas.  That is we shall take $V_2$ to
be negligible at any one time but nevertheless to be present to
perform its role of redistributing hyper-angular-momentum.  Then by
(3.18) our `energy' equation takes the form
$$\sum {\bf Q}'^2_I = {\cal L}^2 \ , $$
where on the right we have a conserved quantity.

As far as the statistical mechanics is concerned the ${\bf Q}_I$ are
equivalent, so we invent a 6 dimensional phase space ${\bf Q}, {\bf
Q}'$ and divide it into cells of equal volume.  The number of ways $W$
of putting $n_a$ distinguishable particles in the $a^{{\rm th}}$ cell
centred on ${\bf Q}_a, {\bf Q}_a'$ is 
$$W = N! \bigg / \left ({\prod _a} n_a!\right) \ .$$
We maximise $W$ subject to the constraints
$$\ \ \ \, \; \! \sum n_a  =  N$$
\vspace{-.2in}
$$\sum n_a {\bf Q}'^2_a  = {\cal L}^2 \eqno (3.21)$$
\vspace{-.1in}
$$ \!  \! \! \sum n_a {\bf Q}^2_a = 1$$ 

and using Lagrange multipliers ${\tilde {\alpha}}, {\textstyle {1 \over
2}} {\tilde \beta}, {\textstyle {1 \over
2}} {\tilde \gamma}$ for the constraints we obtain
$$\delta {\rm ln}W - \sum \delta  n_a (\tilde \alpha + {\textstyle {1 \over 2}}
\tilde \beta {\bf Q}'^2_a + {\textstyle {1 \over 2}} \tilde \gamma
{\bf Q}^2_a) = 0 \ .$$  
The normal use of Stirling's theorem then leads to 
$$n_a = \exp - \left[ \tilde \alpha - 1 + \tilde \beta {\textstyle {1 \over
2}} {\bf Q}'^2_a + \tilde \gamma {\textstyle {1 \over 2}} {\bf Q}^2_a \right] \
. \eqno (3.22)$$

Returning to the constraint equations (3.21) we see that they are the
same with ${\bf Q}'^2_a / {\cal L}^2$ and ${\bf Q}^2_a$ interchanged.
Hence when we solve them for $\tilde \beta {\cal L}^2$ and $\tilde
\gamma$ we inevitably find those to be equal.  If we now pass to a
continuous distribution function $f ({\bf Q}, {\bf Q}')$ in our
six-dimensional phase space by replacing $n_a$ by $f({\bf Q}, {\bf
Q}') d^3 Q\, d^3 Q'$ we find
$$f ({\bf Q}, {\bf Q}') = A \exp [ - \tilde \beta {\textstyle {1 \over
2}} ({\bf Q}'^2 + {\cal L}^2 {\bf Q}^2) ] \ . \eqno (3.23)$$
Where ${\bf Q}'^2$ stands for any one of the equivalent ${\bf Q}'^2_I$
and ${\bf Q}^2$ stands for one of the equivalent ${\bf Q}^2_I = {m_I
\over M} q^2_I/r^2$.  We shall now re-express the ${\bf Q}'_I$ in terms of
the ${\bf q}_I$ and the ${\dot {\bf q}}_I$
$${\bf Q}'^2_I = {m_I \over M} \left( {{\bf q}_I \over r} \right)'^2 =
{r^4 \over M} \, m_I \left[ {d \over dt} \left( {{\bf q}_I \over r}
\right) \right] ^2 = {r^2 \over M}\, m_I \left( {\dot {\bf q}}_I - {{\dot
r} \over r} {\bf q}_I \right)^2 \ . $$

Inserting these expressions into (3.23) and writing $H$ for the
`Hubble' expansion rate ${\dot r} / r$, which is of course time
dependent in our problem, the expression becomes simplest written in
terms of the peculiar velocity ${\bf v}_I$ relative to the Hubble flow
$${\bf v}_I = {\dot {\bf q}}_I - H {\bf q}_I \ . \eqno (3.24)$$
The distribution of ${\bf v}_I$ and ${\bf q}_I$ at given $r$ is then
$$f ({\bf v}_I, {\bf q}_I|r) = A \, \exp \left[ - {{\tilde \beta} r^2
\over M} \left( {\textstyle {1 \over 2}} m_I {\bf v}^2_I \right) -
{{\tilde \beta} {\cal L}^2 \over r^2} \left( {\textstyle {1 \over 2}}
{m_I \over M} {\bf q}^2_I \right) \right] \ . \eqno (3.25)$$ $r$ is
continually changing in time and by no means slowly, nevertheless
${\tilde \beta}$ and ${\cal L}^2$ are exactly constant however slowly
the interactions have led to the equilibrium distribution.  We notice
that at each $r$ the peculiar velocity ${\bf v}_I$ with respect to the
`Hubble' flow $H {\bf q}_I$ is Maxwellianly distributed with a
temperature proportional to $r^{-2}$ and the heavier particles move
more slowly with respect to that flow.  Likewise $m_I^{1/2} {\bf q}_I$
are Gaussianly distributed with a dispersion proportional to $r$.  The
value of ${\tilde \beta}$ is readily deduced from the condition $\sum
{m_I \over M} {\bf q}^2_I = r^2$ which gives ${\tilde \beta} =
(3N$--1$)/{\cal L}^2$.  By (3.23) $f$ is unchanging so there is no
entropy change. Here the $-1$ follows from the fact that there are
$3N$--1 independent ${\bf Q}_I$ components when the constraint (3.20) is
imposed.  Had we imposed also the barycentre constraints we would have
found $3N$--4.  The sharing of ${\cal L}^2$ without the sharing of
energy has arisen before in Cometary Theory (see Rauch \& Tremaine,
1996).

\section{Conclusions}

It is natural to ask whether the new form of equilibrium (3.25) can be
generalised to the Hubble flow of relativistic cosmology and to look
for connections with the interesting fact that the cosmic background
radiation remains in equilibrium during cosmic expansion and creates
no entropy while a classical gas creates entropy during uniform
expansion due to its bulk viscosity.  These are beyond the scope of
this paper but we hope they show that exploration of somewhat esoteric
N-body problems can have considerable interest outside classical
dynamics.  Readers will have appreciated the beauty of these N-body
problems treated exactly in $3N$ dimensions as well as the novelty of
the unusual statistical mechanics of systems with isotropic expansions
as seen in the distribution (3.25) where the velocities are those
relative to the time-dependent flow as given by (3.24).

\begin{appendix}

\section{}

Motion in Henon's (1959) isochrone potential $V = -GM^2/(b+s)$ where $s^2 = r^2 +
b^2$.  We write $\epsilon = E/M$ for the specific energy.  We need to
evaluate the integral (2.13) for $t$ and (2.15) for $\phi$.  If we write
$$a = GM/(-2 \epsilon) \ , \eqno ({\rm A}1)$$ and put $$(1-e^2) =
L^2/[-2 \epsilon (a-b)^2] \ , \eqno ({\rm A}2)$$ then the surd in (2.13)
$\times r$ is $$S = [2 \epsilon (s^2 - b^2) + 2GM (s-b) - L^2]^{1/2} =
(-2 \epsilon)^{1/2} [(a-b)^2 e^2 - (s-a)^2]^{1/2} \ ; \eqno ({\rm
A}3)$$ so we write $$(s-a) = - (a-b) e \cos \eta \ , \eqno ({\rm A}4)$$ and
(2.13) becomes $$t-t_0 = \int (s/S) ds = a (-2 \epsilon)^{-1/2} (\eta -
(1 - b/a) e \sin \eta) \ . \eqno ({\rm A}5)$$ Putting $$\kappa = (-2
\epsilon)^{1/2}/a = (GM/a^3)^{1/2} = (-2 \epsilon)^{3/2}/GM \ , \eqno
({\rm A}6)$$ then $2 \pi/\kappa$ is the radial period as for Kepler's
case and
$$\kappa (t-t_0) = \eta - (1 - b/a) e \sin \eta \ ,  \eqno ({\rm A}7)$$
which we recognise as the generalisation of Kepler's equation $(b =
0)$.  Both in Kepler's case and for the isochrone, direct evaluation
of (2.15) in terms of $\eta$ gives untidy formulae (cf. (A17)).  In the
Keplerian case one uses $1/r$ as the integration variable to get
$\ell/r = 1 + e \cos \phi$.  For the isochrone one has from (2.15)
$$\phi = {\textstyle {1 \over 2}} L \int \left( {1 \over s - b} + {1
\over s + b} \right) {ds \over S} \ , \eqno ({\rm A}8)$$ and we need to
use ${1 \over s-b}$ {\it and} ${1 \over s+b}$ instead of $1/r$.  We
therefore use two new angles $\chi$ and $\chi_{_+}$ and rewrite (A4) in
terms of $a_p$ the pericentric value of $s-b$.
$$\sqrt{r^2 + b^2} - b = s - b = {a_p \over 1-e} (1 - e \cos \eta) =
{a_p (1+e) \over 1 + e \cos \chi} = -2b + {(a_p + 2b)(1 + e_{_+}) \over
(1 + e_{_+} \cos \chi_{_+})} \eqno ({\rm A}9)$$ where $$a_p = (a-b) (1-e) \ , 
\eqno ({\rm A}10)$$ and, taking apocentre values and writing $$f = {2b
\over a_p} \ , \eqno ({\rm A}11)$$ we find $${1+f \over 1 - e_{_+}} = {1 \over 2} (1+f) +
{1 \over 1-e} \ . \eqno ({\rm A}12)$$
The two parts of the integral (A8) then give $$\phi = {\textstyle {1
\over 2}} \left[ \chi + {L \over \sqrt{L^2 + 4GMb}} \chi_{_+} \right] \
. \eqno ({\rm A}13)$$  Since both $\chi$ and $\chi_{_+}$, (and $\eta$)
increase by $2 \pi$ in one radial period we find that $\phi$ increases
by $$\Phi = \pi \left[ 1 + {L \over \sqrt{L^2 + 4 GMb}} \right] < 2
\pi \ . \eqno ({\rm A}14)$$  Only when $b = 0$ is $\Phi = 2 \pi$ and
then both $\chi$ and $\chi_{_+}$ reduce to $\phi$.  Thus the orbits do
not close (unless ${L \over \sqrt{L^2 + 4 GMb}}$ is rational) and they
form rosettes with inner and outer radii of $$r_p = \sqrt{a_p (a_p
+2b)} \ \ {\rm and} \ \ r_a = \sqrt{a_a (a_a + 2b)} \ , \eqno ({\rm
A}15)$$ where $$a_a = a_p (1+e)/(1-e) = (a - b) (1 +e) \ . \eqno ({\rm
A}16)$$

If in place of inventing $\chi$ and $\chi_{_+}$ we proceed with (A4), we
obtain the ugly formula (cf. (2.36K))
$$\phi = \tan^{-1} \left( \sqrt{{1 - e \over 1 + e}} \tan \eta/2
\right) + {L \over \sqrt{L^2 + 4 GMb}} \, 
\tan^{-1} \left( \sqrt{{1 - e_{_+} \over 1 + e_{_+}}} \tan \eta/2 \right) \ , \eqno ({\rm A}17)$$ which is term for term the
same as (A13).

One may show from the above formulae that
$${L \over \sqrt{L^2 + 4 GMb}} = \sqrt{{1 - e^2 \over 1 - e^2_+}}\, {e_{_+}
\over e} \ . \eqno ({\rm A}18)$$
To get the general solution to the N-body problem one procedure is to
take a value of $\eta$, from that determine $t$ from (A5), $s$ from
(A4) and $r = \sqrt{s^2-b^2}$ and finally $\phi$ from (A17).  We
finally have $${\bf r} = r ({\bf A} \sin \phi + {\bf B} \cos \phi)$$
as before.  (A9) gives the basic relationship between $r, s, \eta,
\chi$ and $\chi_{_+}$ whereas (A13) relates $\phi$ to $\chi$ and $\chi_{_+}$
and (A5) relates $t$ to $\eta$.
\end{appendix}

\section*{Note Added in Proof}

Geometrical Insight gives greater clarity.  From (2.5), the problem is
a central one both in 3$N$ dimensions and for each particle.  Together
the initial ${\bf r}$ and ${\dot {\bf r}}$ define a plane in 3$N$
dimensions through the centre.  The central force lies in that plane
so all the motion continues in it. The motion is just planar motion
under the potential $V(r)$ so we get the usual orbit.  The orbit of
each particle is the projection of this 3$N$ planar orbit onto the
orbital plane of that particle (that has most of the 3$N$ coordinates
fixed at zero).  Thus in all cases the particle orbits are projections
of the 3$N$ planar orbit which is a familiar one.  If this is an
ellipse with the barycentre as its focus, then the particle orbit will
be the projected ellipse but the projection does not preserve the
barycentre as its focus.  Similarly if the 3$N$ orbit is a rosette,
the individual particle orbits will be projected rosettes, i.e.,
rosettes between similar ellipses.  That will happen for most
potentials, e.g., the isochrone.

\label{lastpage}

\end{document}